\documentclass[journal]{IEEEtran}

\usepackage{amsmath,amsfonts,amssymb}
\usepackage{graphicx}
\usepackage{cite}
\usepackage{multirow}
\usepackage{algorithm}
\usepackage{algorithmic}
\usepackage{subfigure}
\usepackage{url}
\usepackage{color}
\scrollmode
\IEEEoverridecommandlockouts

\newtheorem{example}{Example}

\begin{document}
\title{Reliable Broadcast to A User Group with Limited Source Transmissions}
%\title{A Linear Network Coding Solution for double-unicast network}
\author{\authorblockN{Xiaoli~Xu, Meenakshi Sundaram Gandhi Praveen Kumar, and Yong~Liang~Guan\\}
\authorblockA{School of Electrical and Electronic Engineering, Nanyang Technological University, Singapore\\
\{xuxiaoli, meenakshi, eylguan\}@ntu.edu.sg}}
%\thanks{X. Xu and Y. L. Guan are with the School of Electrical and Electronic Engineering, Nanyang Technological University, Singapore 639801
%(email: \{xuxiaoli, eylguan\}@ntu.edu.sg)}
%\thanks{Y. Zeng was with the School of Electrical and Electronic Engineering, Nanyang Technological University, Singapore. He is now with the Department of Electrical and Computer Engineering, National University of Singapore. (email: elezeng@nus.edu.sg)}
%\thanks{T. Ho is with the Department of Electrical Engineering, California Institute of Technology, Pasadena, California 91125, USA (email: tho@caltech.edu)}
%\thanks{This work was supported by the Advanced Communications Research Program DSOCL06271, a research grant from the Directorate of Research and Technology (DRTech), Ministry of Defence, Singapore.}

\maketitle

\begin{abstract}
In order to reduce the number of retransmissions and save power for the source node,  we propose a two-phase coded scheme to achieve reliable broadcast from the source to a group of users with minimal source transmissions.  In the first phase, the information packets are encoded with batched sparse (BATS) code, which are then broadcasted by the source node until the file can be \emph{cooperatively} decoded by the user group. In the second phase, each user broadcasts the re-encoded packets to its peers based on their respective received packets from the first phase, so that the file can be decoded by each individual user.  The performance of the proposed scheme is analyzed and the rank distribution at the moment of decoding is derived, which is used as input for designing the optimal BATS code. Simulation results show that the proposed scheme can reduce the total number of retransmissions compared with the traditional single-phase broadcast with optimal erasure codes. Furthermore, since a large number of transmissions are shifted from the source node to the users,  power consumptions at the source node is significantly reduced.
\end{abstract}

\section{Introduction}
Wireless broadcast models the communication scenario where common information is to be transmitted from the source to a set of receivers through wireless channels. It has a wide range of applications, such as satellite communication and file distribution. However, in wireless communications, some information may be lost due to deep fading. To achieve reliable broadcast, retransmission or erasure coding \cite{Luby2006} has been applied. Traditional ``Repeat Request-retransmission" is bandwidth-inefficient when there are many receivers. On the other hand, the erasure codes are more efficient in terms of bandwidth occupation, but they usually require high encoding/decoding complexity and incur significant delay when the file size is large. Network coding has been considered as a promising strategy to improve the efficiency of the retransmission schemes \cite{Nguyen2008,Fragouli2006}. However, this scheme requires prompt and correct feedbacks from all the receivers.

All the aforementioned broadcast schemes assume that there is no cooperations among the receivers, and hence reliable broadcast can only be achieved by source retransmissions, either with or without coding. In this paper, we consider the scenario where the source node is intended to broadcast some common information to a group of users that are closely located within a small region. Such a setup models various practical communication scenarios, e.g., video streaming from the base station to a group of nearby  mobile users in cellular networks, the communication from an isolated user to a squadron of users in ad-hoc networks, etc.  In this case, it is desirable to minimize the number of retransmissions by the source node to save its power/bandwidth. To this end, we propose a two-phase transmission scheme, with source broadcasting in the first phase and peer-to-peer (P2P) packet exchange in the second phase. Specifically, in the first phase, the information packets are  broadcasted by the source node until the file can be \emph{cooperatively} decoded by the user group. In the second phase, the users exchange the re-encoded packets to its peers based on their respective received packets from the first phase, so that the file can be decoded by each individual user with minimum number of retransmissions.

P2P erasure repair for wireless video broadcasting was firstly studied in \cite{Sanigepalli2006} for Multimedia Broadcast/Multicast Service (MBMS) applications, where various suboptimal scheduling schemes are proposed.  Give the global state information of all users, the cooperative P2P repair (CPR) problem was formulated and proved to be NP-hard in \cite{Cheung2006}. By assuming optimal inter-user channels in the second phase, a distributed CPR algorithm was proposed in \cite{CPR}. However, the proposed scheme still heavily relies on the perfect control information exchange and the efficiency of the retransmission in this scheme is relatively low.  Then, the CPR algorithm in \cite{CPR} was extended to network coded CPR (NC-CPR) in \cite{DCPR} by applying random linear network coding\cite{Ho06} in P2P communications.  By integrating the idea of NC-CPR \cite{DCPR} and rarest first scheduling, a light-weight peer scheduling algorithm was proposed in \cite{Fan2010}.

All the existing schemes discussed above \cite{Sanigepalli2006,Cheung2006,CPR,DCPR,Fan2010} assume that the inter-user channels used in the second phase are lossless. Consequently, some state information can be reliably exchanged before the start of the P2P communications. Their performance may degrade severely if the state/control information is lost. In this paper, we propose a fully distributed two-phase cooperative broadcast scheme that does not require exchange of any state information, and hence can be applicable for networks with lossy channels. Furthermore, the existing schemes focus on the scheduling algorithms of the second phase, with the assumption that the users can recover the file cooperatively based on their received packets in the first phase. However, such assumption is not guaranteed in wireless networks, e.g., for a network with $k$ users, each experiencing erasure probability $p$,  a packet sent by the source node is lost forever with a non-zero probability $p^k$. The residual loss is considered bearable in video streaming, but it may not be acceptable in file distribution. To guarantee the file recovery, the design of the proposed scheme involves both phases.

First, we propose to apply good erasure code, such as fountain code, on the information packets to be transmitted. The source stop transmission after the user group as a whole has received a sufficient number of coded packets.  At the end of the first phase, the user group, if allowed to decode cooperatively,  should be able to recover the file, although each individual user still cannot decode the information based on their own respective received packets. Therefore, in the second phase, the users help each other by broadcasting to its peers based on their respective received packets via the P2P communication network \cite{Cohen2003}.  Random linear network coding (RLNC)  has been shown to significantly improve the bandwidth efficiency in P2P communications. However,  network coding overhead increases linearly with the number of packets coded together. To maintain the efficiency, a file is usually divided into segments and network coding is only performed within each segment. As an integration of the erasure code in the first phase and network code in the second phase,  batched sparse (BATS) code  \cite{Yang11} is adopted in the proposed two-phase broadcast scheme.

Batch sparse (BATS) code was proposed in \cite{Yang11} for improving the throughput of wireless erasure networks.  The performance of BATS code is largely dependent on the predefined degree distribution which is optimized based on the channel rank distribution \cite{Yang13}. In wireless erasure networks with fixed topology, the channel rank distribution can be obtained based on erasure probability of each link. However, in P2P networks, the channel rank distribution is also affected by the communication protocols. For example, consider two batches, $\mathcal{B}_i$ and $\mathcal{B}_j$, if more packets are transmitted for $\mathcal{B}_i$ than $\mathcal{B}_j$ in the second phase, the rank of $\mathcal{B}_i$ is likely to be larger than the rank of $\mathcal{B}_j$ at the moment of decoding. Furthermore, since the batch size is $M$, the source node only sends $M$ packets for each batch in the first phase. In the second phase, the rank of $\mathcal B_i$ should not exceed the total number of packets received by the user group in the first phase, which is smaller or equal to $M$. In this paper, we propose a distributed algorithm for each user to determine the transmission sequence and analyze the channel rank distribution at the moment of decoding. The analytical channel rank distribution is shown to match very well with the simulation results, and the BATS code generated based on the analytical rank distributions can be decoded at a small overhead. Compared with traditional broadcast based on repetitive retransmission or erasure codes, the proposed two-phase scheme achieves reliable broadcast with less number of transmissions. Moreover, a large amount of transmissions are shifted from the source to the users, where much less power is required per transmission.

\emph{Notations}: Throughout this paper, random variables are represented by boldface upper-case letters and the probability of an event is presented by $\Pr(\cdot)$. For a random variable $\mathbf X$, we use $\mathbb{E}[\mathbf X]$ to denote the expectation of $\mathbf X$. Furthermore, scalars, vectors and matrices are represented by italic, boldface lower- and boldface upper-case letters, respectively.

\section{System Model}\label{sec:model}
\begin{figure}[htb]
\centering
\includegraphics[scale=0.7]{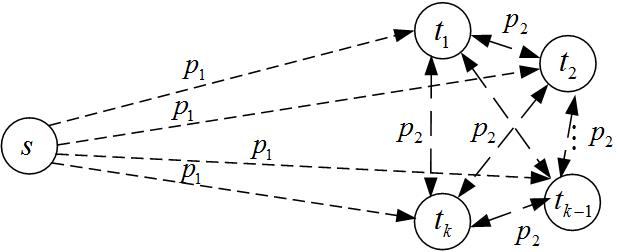}
\caption{A source broadcasts to a group of users.}
\label{F:model}
\end{figure}

We consider the problem where a source node $s$ wants to broadcast common information to a group of $k$ users, which are closely located within a small region far away from the source node, as shown in Fig.~\ref{F:model}.    Assume that the link between the source to each user is an independent\footnote{The channels are assumed to be independent when the distance between any pair of users is larger than half of the wavelength.} memoryless channel with erasure probability $p_1$ and the links between the users have independent erasure probability $p_2$, where $0<p_1,p_2<1$.  Since the distance from the source to the user group is much larger than that between the users,  we have $p_2<p_1$.

As the channel between the source and the user group is less reliable and much more power is required to compensate for the path loss over the large transmission distance from the source to the users, it is desirable to limit the number of transmissions by the source node; instead, the users can help each other by exchanging their respective received packets through the more reliable inter-user channels, to ensure that every user is able to recover the file.

\section{A Two-Phase Protocol with BATS Code}\label{sec:main}
In this section, we propose a two-phase transmission protocol based on BATS code to achieve reliable communication in Fig.~\ref{F:model}. A BATS code contains an outer code and an inner code. The outer code is an extension into the matrix form of the traditional fountain code. Specifically, the source node first obtains a degree $d$ by sampling a predefined degree distribution, and then randomly picks $d$ distinct input packets to generate a batch of $M$ fountain-coded packets. The batches are transmitted sequentially. The inner code of BATS employs RLNC at the intermediate nodes and only packets within the same batch are coded together. Hence, the coding overhead is limited by batch size $M$, which is usually considered as negligible compared with the packet length. Finally, the inner and outer codes are jointly decoded at the receiver with belief-propagation (BP) decoding.

Based on BATS code, we propose to achieve reliable broadcast in Fig.~\ref{F:model} via the following two-phase transmissions:\\
\textbf{Phase 1}: The file is divided into $F$ packets, which are encoded with BATS code of batch size $M$ at the source node. The batches are sequentially sent to the users until a stop criterion is satisfied.\\
\textbf{Phase 2}: The users help each other by exchanging their respective received packets via network coded peer-to-peer (P2P) transmissions, until all the users can recover the file.\\

\subsection{Phase 1 Transmission}\label{sec:phase1}
With optimal BATS code \cite{Yang13}, the file can be recovered if $(1+\eta)F$ packets are received, where $\eta\ll 1$ for moderate to large $F$. Denote by $\mathbf X$ the number of packets received by the user group.  To ensure that the user group can finally recover the file via P2P transmissions with probability no smaller than $(1-\varepsilon)$, we must have
\begin{align}
\Pr(\mathbf X\geq(1+\eta)F)\geq 1-\varepsilon. \label{eq:prob}
\end{align}
Since the channels from the source node to different users are independent, the probability that a packet is received by at least one of the user equals to $(1-p_1^k)$. After $n$ batches, or $nM$ packets, have been sent by the source node, the number of received packets $\mathbf X$ is a random number following binomial distribution $\mathcal{B}(nM,1-p_1^k)$. As $nM$ is usually  large, this binomial distribution can be approximated by the normal distribution $\mathcal{N}(\mu,\sigma^2)$, where $\mu=nM(1-p_1^k)$ and $\sigma^2=nM(1-p_1^k)p_1^k$. Hence, \eqref{eq:prob} can be equivalently written as
\begin{align}
Q\left(\frac{(1+\eta)F-\mu}{\sigma}\right)\geq 1-\varepsilon,\label{eq:thre}
\end{align}
where $Q(x)=\int_{x}^{\infty}\frac{1}{\sqrt{2\pi}}e^{-\frac{\tau^2}{2}}d\tau$ denotes the Gaussian $Q$ function. From \eqref{eq:thre}, we can obtain the minimum value for $n$ as
\begin{align}
n&\geq\frac{F'}{M(1-p_1^k)}-\frac{\alpha}{2M(1-p_1^k)}\left[\sqrt{4p_1^kF'+\alpha^2p_1^{2k}}-\alpha p_1^k\right]\nonumber\\
&\approx \frac{F'}{M(1-p_1^k)}-\frac{\alpha\sqrt{4p_1^kF'}}{2M(1-p_1^k)},
\end{align}
where $F'=(1+\eta)F$ and $\alpha=Q^{-1}(1-\varepsilon)$. Hence, we can set $n$ as
\begin{align}
n=\left\lceil\frac{F'}{M(1-p_1^k)}-\frac{\alpha\sqrt{4p_1^kF'}}{2M(1-p_1^k)}\right\rceil. \label{eq:nValue}
\end{align}
After sending out $n$ batches, the source node stops transmission. Since the expected number of packets received by each user is only $nM(1-p_1)$, which is smaller than $(1+\eta)F$, the file cannot be recovered by each individual user. In the second phase, the users help each other  by broadcasting the network coded packets generated from their respective received packets to ensure that all users can successfully recover the file.

%The P2P communication in the second phase is discussed for two cases, distinguished by whether a user can learn the status of the rest $k-1$ users at the end of phase 1.

\subsection{Phase 2 Transmission}\label{sec:blind}
Since all the users are physically distributed, they have no knowledge on  what packets have been received by others at the end of Phase 1. Therefore, during Phase 2 transmission, each user can only determine what packets to be sent based on its own received packets.

Denote by $\mathcal{N}_i^j$ the set of packets received by user $j$ with batch index $i$, where $i\in\{1,...,n\},j\in\{1,...,k\}$.  Consider two typical users $t_j$ and $t_{j'}$.  With random linear network coding, a coded packet generated from $t_j$ for batch $i$ is useful for user $t_{j'}$ if $\mathcal{N}_i^j\setminus \mathcal{N}_i^{j'}\neq\emptyset$. Furthermore, if $|\mathcal{N}_i^j\setminus \mathcal{N}_i^{j'}|=m$, $m$ useful packets for batch $i$ can be generated from $t_j$. Without knowing $\mathcal{N}_i^{j'}$, user $t_j$ can estimate the value of $|\mathcal{N}_i^j\setminus \mathcal{N}_i^{j'}|$ based on its own received packets, $\mathcal{N}_i^j$. Specifically,  $|\mathcal{N}_i^j\setminus \mathcal{N}_i^{j'}|=m$ if $m$ out of $|\mathcal{N}_i^j|$ received packets at user $t_j$ are erased at user $t_{j'}$, i.e.,
\begin{align}
&\Pr\left(|\mathcal{N}_i^j\setminus \mathcal{N}_i^{j'}|=m \Big |\mathcal{N}_i^j\right)\nonumber\\
&\quad =
\begin{cases}
{|\mathcal{N}_i^j|\choose m}p_1^m(1-p_1)^{(|\mathcal{N}_i^j|-m)}, & m\leq |\mathcal{N}_i^j|\\
0, & |\mathcal{N}_i^j|<m\leq M
\end{cases}
\end{align}

Assume that user $t_j$ has already sent out $u$ packets generated from batch $i$. Then, the $(u+1)$th packet generated from the same batch is still useful for user $t_{j'}$ if either i) $m\geq u+1$, or ii) at most $(m-1)$ out of $u$ packets are received by $t_{j'}$. Denote this event by $E_i^u(j)$, whose probability can be estimated as
{\small
\begin{align}
&\Pr(E_i^u(j)|\mathcal{N}_i^j)\nonumber\\
&=\sum_{m=u+1}^{M}\Pr(m|\mathcal{N}_i^j)+\sum_{m=1}^{u}\Pr(m|\mathcal{N}_i^j)\sum_{l=0}^{m-1}{u\choose l}(1-p_2)^lp_2^{(u-l)}.\label{eq:estimation}
\end{align}}

By symmetry, \eqref{eq:estimation}  applies for all other peers of user $t_j$. Hence, $\Pr(E_i^u(j)|\mathcal{N}_i^j)$ can be used as a metric to measure the usefulness for user $t_j$ to broadcast $(u+1)$th packet generated from batch $i$.   In general, \eqref{eq:estimation} is valid for any  $u\geq 0$. However, a user usually will not send more than $M$ packets for one batch before the moment of decoding, and hence we may only calculate the estimation up to $u=M$. Denote  $\mathbf{S}_j\in\mathbb{R}^{M\times n}$ for $j$ with the $(u,i)$th element equal to $\Pr(E_i^u(j)|\mathcal{N}_i^j)$. As more packets are sent out from a batch, the new transmission is likely to be less useful. Hence, each column of $\mathbf{S}_j$ is a monotonically decreasing vector. Furthermore, if more packets are received for batch $i$ than batch $i'$ at the end of Phase 1,  the packet generated from batch $i$ is more likely to be useful than that from batch $i'$. Mathematically, if $|\mathcal{N}_{i}^{j}|>|\mathcal{N}_{i'}^{j}|$, we have $\Pr(E_i^u(j)|\mathcal{N}_i^j)> \Pr(E_{i'}^u(j)|\mathcal{N}_{i'}^j)$ for any $u\geq 0$.

To maximize the spectral efficiency, a packet that is expected to be more useful should be transmitted with higher priority.  To obtain the optimal transmission order, user $t_j$ sorts all the elements in $\mathbf{S}_j$ in descending order. Denote the ordered elements by a vector $\mathbf{s}_j\in\mathbb{R}^{1\times Mn}$ and  the column index of the ordered elements by a vector $\mathbf v_j\in\mathbb{Z}^{1\times Mn}$. Then each element of $\mathbf{v}_j$ represents a batch ID within $\{1,...,n\}$. User $t_j$ sequentially transmit the coded packets with batch ID obtained from $\mathbf v_j$.

\begin{example}
For  illustration purpose, we consider a simple setup with $p_1=0.5$, $p_2=0.1$, $M=4$ and $n=5$. If user $t_j$ receives $\{2,1,3,4,2\}$ packets at the end of Phase 1 for batch $1$ to $5$, respectively, the usefulness matrix calculated from \eqref{eq:estimation} is
\begin{align}
\mathbf{S}_j=\left[
\begin{matrix}
0.7500   & 0.5000  &  0.8750 &   0.9375  &  0.7500\\
0.3000   & 0.0500  &  0.5375  &  0.7125  &  0.3000\\
0.0525  &  0.0050  &  0.2000   & 0.3862 &   0.0525\\
0.0075  &  0.0005  &  0.0448  &  0.1410  &  0.0075\\
\end{matrix}
\right].
\end{align}
By sorting all the elements in $\mathbf{S}_j$, user $t_j$ can obtain $\mathbf{s}_j=\left[\begin{matrix}0.9375 & 0.8750 & 0.7500 & 0.7500 & 0.7125 & \cdots \end{matrix}\right]$. The column indices of the elements in $\mathbf{s}_j$ give the transmission order as $\mathbf{v}_j=\left[\begin{matrix}4&3 &1 &5 &4 &3 &\cdots\end{matrix}\right]$. At the first time slot when user $t_j$ can access the channel, it will send a coded packet generated from all the available packets for batch 4, received both in Phase 1 and in Phase 2.  For the second transmission,  a coded packet from batch 3 will be sent out and the process continuous until all the users are able to decode the file.
\end{example}

%Ideally, user $t_j$ should update $\mathbf{S}_j$ when it receives any packet from its peers.
\subsubsection{Estimate the Total Number of Transmissions in Phase 2}
We assume that all the users have equal probability for transmission under a multiple-access scheme, such as TDMA or CSMA/CA. The transmission stops when all the users are able to recover the file, which is assumed to happen after $T$ transmissions in total. On average, user $t_j$ will send out $\frac{T}{k}$ coded packets with batch ID given by the first $\frac{T}{k}$  elements of $\mathbf v_j$. Among all the $\frac{T}{k}$ packets sent out by $t_j$, only $\frac{(1-p_2)T}{k}$ will reach user $t_{j'}$ due to packet erasures. Hence, the total number of packets received by user $t_{j'}$ from its $k-1$ peers is a function of $T$, which is $P(T)=\frac{(1-p_2)(k-1)T}{k}$. By symmetry, we may assume that the batch IDs of the received packets follow uniform distribution. In other words, if we denote by $\mathbf Y_2$ the total number of packets received for a typical batch in Phase 2, then $\mathbf{Y}_2$ is a random number following binomial distribution $\mathcal{B}(P(T),\frac{1}{n})$, i.e.,
\begin{align}
&\Pr(\mathbf{Y}_2=i)={P(T)\choose i}\left(\frac{1}{n}\right)^i\left(1-\frac{1}{n}\right)^{P(T)-i},\nonumber
\\ &\quad i=0,...,P(T). \label{eq:PY2}
\end{align}

When $P(T)$ is much larger than $n$, we may assume that on average $\frac{P(T)}{n}$ packets are received for each batch. Moreover, denote by $\mathbf{Y}_1$ and $\mathbf{Z}$ the number of packets for this batch received by a single user and the  user group in Phase 1, respectively. Clearly, we have $\mathbf{Y}_1\sim\mathcal{B}(M,1-p_1)$, i.e.,
\begin{align}
\Pr(\mathbf{Y}_1=i)={M\choose i}(1-p_1)^{i}p_1^{(M-i)}, i=0,...,M.\label{eq:PY1}
\end{align}
Furthermore, $\mathbf Z$ is a random variable which satisfies $\mathbf{Z}\geq\mathbf{Y}_1$ since the union of all sets is always larger than any single set. The distribution of $\mathbf Z$ given $\mathbf{Y}_1$ can be obtained as
\begin{align}
&\Pr(\mathbf{Z}=j|\mathbf Y_1=i)\nonumber\\
&={{M-i}\choose {j-i}}(1-p_1^{k-1})^{j-i}p_1^{(k-1)(M-j)}, \forall j\geq i, i=0,...,M\label{eq:PZgY1}
\end{align}

Since only $\mathbf Z$ packets are available for the whole user group, the number of received packets by each user from its peers cannot be larger than $\mathbf{Z}$. Any more packets received will be a linear combination of the existing $\mathbf{Z}$ packets. Out these $\mathbf Z$ packets, the user already has $\mathbf{Y}_1$ packets  before Phase 2, which are received during Phase 1. Hence,  any more than $(\mathbf{Z}-\mathbf{Y}_1)$ packets received during Phase 2 will be redundant.  Furthermore, since random linear network coding over a sufficiently large field size is applied, we assume that any packets received before reaching its limit $\mathbf Z$ is innovative. Clearly, the larger $T$ is, the more redundant packets will be received. Therefore, the expected number of redundant packets for all batches received during phase 2, denoted by $R(T)$, is a function of $T$, which can be estimated as:
{\small
\begin{align}
&R(T)\nonumber\\
&=n\sum_{i=0}^M\sum_{j=i}^M \left(\frac{P(T)}{n}-j+i\right)^{+}\Pr(\mathbf{Z}=j|\mathbf{Y}_1=i)\Pr(\mathbf Y_1=i)\label{eq:redun},
\end{align}}
where $(x)^+\triangleq\max\{0,x\}$ and  $\Pr(\mathbf{Y}_1)$ and $\Pr(\mathbf Z|\mathbf Y_1)$ are given in \eqref{eq:PY1} and \eqref{eq:PZgY1}, respectively.

%\begin{align}
%R(T)&=n\sum_{i=0}^M\sum_{j=i}^M\sum_{l=j-i}^{P(T)}(l-j+i)\Pr(\mathbf{Y}_1=i,\mathbf{Z}=j,\mathbf{Y}_2=l)\nonumber\\
%&\overset{(a)}{=}n\sum_{i=0}^M\sum_{j=i}^M\sum_{l=j-i}^{P(T)}(l-j+i)\Pr(\mathbf{Z}=j|\mathbf{Y}_1=i)\Pr(\mathbf Y_1=i)\Pr(\mathbf{Y}_2=l)\label{eq:redun},
%\end{align}
%where $\Pr(\mathbf{Y}_2)$, $\Pr(\mathbf{Y}_1)$ and $\Pr(\mathbf Z|\mathbf Y_1)$ are given in \eqref{eq:PY2},\eqref{eq:PY1} and \eqref{eq:PZgY1}, respectively.

At the end of the transmission, the total number of innovative packets received by user $t_{j'}$ from both phase 1 and phase 2  is $(1-p_1)nM+P(T)-R(T)$. Since decoding the BATS code requires $(1+\eta)F$ packets, we have
\begin{align}
(1-p_1)nM+P(T)-R(T)>(1+\eta)F,\label{eq:determineT}
\end{align}
based on which we can solve the value for $T$.

\subsubsection{Estimate the Rank Distribution}
In conventional directed \emph{acyclic} network considered in \cite{Yang13}, the rank distribution of the batches is determined by the network topology and the erasure probability of each link. However, for the network shown in Fig.~\ref{F:model} with cycles, the rank distribution is affected by the scheduling scheme in Phase 2 transmission.  Since rank distribution is an important parameter for designing BATS code, it is crucial to get a good estimation of the rank distribution before transmission, which is pursued in this subsection.

With given $T$ obtained from the preceding subsection, we are now ready to estimate the rank distribution.  Following similar assumptions as in \eqref{eq:redun}, the rank of a typical batch for a user is $r$ if either of the following two events occur: i) the user group has more than $r$ packets for this batch, but this user only receive $r$, i.e., $\mathbf Z>r$ and $\mathbf{Y}_1+\mathbf{Y}_2=r$; ii) This user receives more than $r$ packets, but only $r$ of the received packets are innovative, i.e., $\mathbf{Y}_1+\mathbf{Y}_2\geq r$ and $\mathbf{Z}=r$.  Hence, at the end of the transmissions, a batch has rank $r, r\in\{0,1,...,M\}$ for a user is given by
\begin{align}
&\Pr(r)=\Pr(\mathbf{Z}>r,\mathbf Y_1+\mathbf Y_2=r)+\Pr(\mathbf{Z}=r,\mathbf Y_1+\mathbf Y_2\geq r)\nonumber\\
&=\sum_{i=0}^{r}\sum_{j=r+1}^{M}\Pr(\mathbf Z=j|\mathbf Y_1=i)\Pr(\mathbf Y_1=i)\Pr(\mathbf{Y}_2=r-i)\nonumber\\
&\quad +\sum_{i=0}^{r}\sum_{j=r-i}^{P}\Pr\left(\mathbf Z=r|\mathbf{Y}_1=i\right)\Pr(\mathbf Y_1=i)\Pr(\mathbf Y_2=j),\label{eq:Rank}
\end{align}
where each individual probability $\Pr(\mathbf{Y}_2)$, $\Pr(\mathbf{Y}_1)$ and $\Pr(\mathbf Z|\mathbf Y_1)$ are given in \eqref{eq:PY2},\eqref{eq:PY1} and \eqref{eq:PZgY1}, respectively.

%The total number of innovative packets received by user $t_{j'}$ from $t_{j}$, denoted by  $Y_{j'}^{j}$, can be estimated as
%\begin{align}
%Y_{j'}^{j}=(1-p_2)\sum_{i=1}^{T/k}\mathbf{s}_{j}(i)\overset{(a)}{\leq} \frac{T(1-p_2)}{k},
%\end{align}
%where $(a)$ follows as $\mathbf{s}_j(i)\leq 1$. When $F$ is relatively large, we may approximate $Y_{j'}^{j}$ by its upper bound $\frac{T(1-p_2)}{k}$ since most of the packets sent out are innovative. Hence, the total number of packets received user $t_{j'}$ may receive $Y_{j'}=\sum_{j\in\{1,...k\}\setminus j'}Y_{j'}^j$

\section{Numerical Results}
Numerical results are provided in this section to corroborate our proposed scheme. We assume that a file of size $2.1$Mb needs to be transmitted  from the source node to a group of 3 users. The erasure probability for the channels between source node to each user is $p_1=0.5$ and the inter-user links are assumed to have erasure probability $p_2=0.1$. The file is divided into $F=2083$ packets, each of length $1000$ bytes.
A BATS code with batch size $M=16$ is applied. The overhead $\eta$ of BATS code is set as $5\%$. To ensure that the file can be recovered with high probability, we set $\varepsilon=10^{-6}$. Based on \eqref{eq:nValue}, the number of batches to be sent out by the source is $n=162$, i.e., the source stops transmission after sending out $162$ batches or $2592$ packets.

\begin{figure}[htb]
\centering
\includegraphics[scale=0.45]{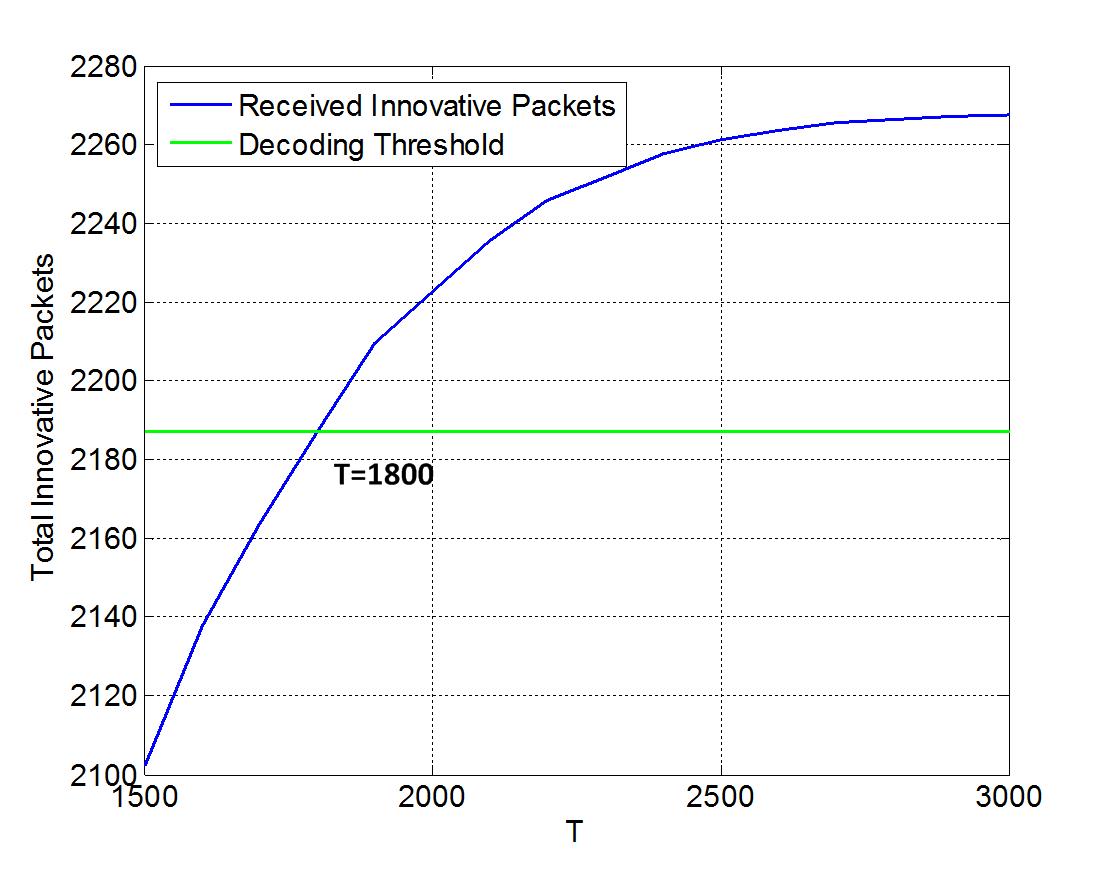}
\caption{Number of innovative packets received by a user versus the total number transmissions in Phase 2.}
\label{F:GetT}
\end{figure}

The total number of innovative packets received by a user is plotted versus the number of transmissions in Phase 2 based on \eqref{eq:determineT} in Fig.~\ref{F:GetT}. From Fig.~\ref{F:GetT}, we can estimate the total number of transmissions in Phase 2 when all the users are able to recover the file, which is $T=1800$ in the setup under consideration. Note that the number of innovative packets received by a user saturate when $T$ becomes sufficiently large, which is expected since the total number of innovative packets received by a user will not exceed the total packets received by the user group at the end of Phase 1.

\begin{figure}[htb]
\centering
\includegraphics[scale=0.55]{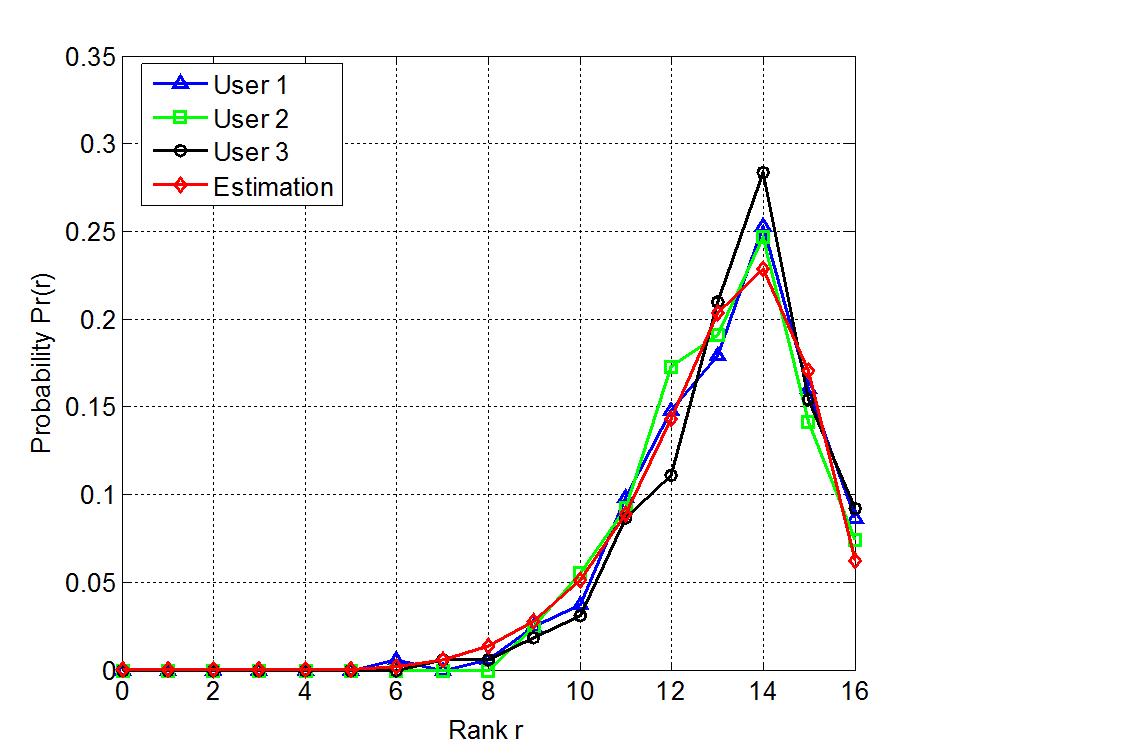}
\caption{Compare the estimated with the simulated rank distributions. }
\label{F:RankDist}
\end{figure}

With $T=1800$, we can estimate the final rank distribution from \eqref{eq:Rank} for $N=0,1,...,M$. The estimated rank distribution is plotted together with the simulated rank distribution for the three users in Fig.~\ref{F:RankDist}. It is observed that the estimated rank distribution matches quite well with the simulation results. Hence, we can design good BATS code based on the rank distribution given in \eqref{eq:Rank}.

\begin{figure}[htb]
\centering
\includegraphics[scale=0.6]{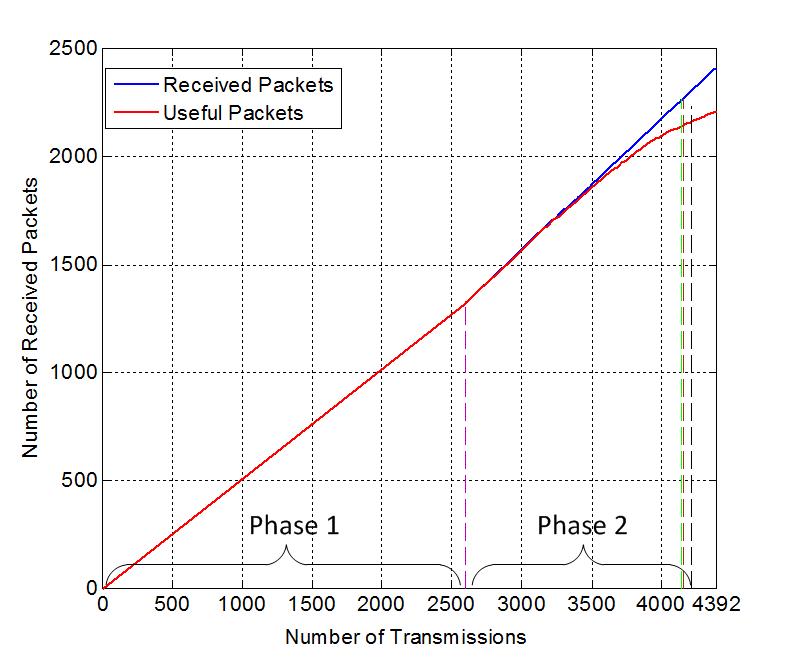}
\caption{The number of received packets versus the number of transmissions for user 1. }
\label{F:simuResult}
\end{figure}

With the rank distribution shown in Fig.~\ref{F:RankDist}, we can get the optimal degree distribution for the BATS code \cite{Yang13}, which is used at the source node. The number of packets received by user $t_1$ is plotted against the number of transmissions in Fig.~\ref{F:simuResult}. At the end of Phase 1, user $t_1$ receives 1317 packets out of 2592 number of packets sent by the source node. As expected, most of packets received by user $t_1$ at the early stage of Phase 2 are useful. The received packets are more likely to be redundant as the number of packets collected by user $t_1$ from its peers increases. Finally, user $t_1$ is able to decode the file when it receives 2272 packets in total, of which 955 packets are from its peers, and it occurs after 1567 P2P transmissions by all the three users. User $t_2$ and user $t_3$ are able to decode to file after 1619 and 1552 P2P transmissions, including the transmissions originated from themselves.

Note that the transmission in Phase 2 stops after 1619 transmissions, which is less than 1800 transmissions, given by the analytical results shown in Fig.~\ref{F:GetT}. This is expected since the overhead for BATS code is usually less than the designed value $5\%$ in practice. However, since the performance of BATS code degrades dramatically for any over-estimation of the channel rank distribution, it is desirable to have a conservative estimation of the channel and hence apply a slightly higher design overhead to guarantee the recovery of the file.

\subsection{Comparison with Single-Phase Transmission}
In this subsection, we compare the proposed two-phase broadcast scheme with the traditional single-phase transmission, where the source keeps transmitting the packets until all the users can recover the file. Assume that optimal erasure code, such as Raptor code \cite{Shokrollahi11}, has been applied so that a user can recover the file after receiving $(1+\eta)F$ packets. Denote by $N$ the minimum number of packets needs to be transmitted by the source so that the file can be recovered by all the users. To find $N$, we further denote by $\mathbf{X}_j$ the number of packets received by user $t_j$, for $j=1,...,k$.  $\mathbf{X}_j$ are independent and identically distributed random variables according to $\mathcal{B}(N,1-p_1)$.  The source transmission stops when all the users are able to recover the file, i.e.,
\begin{align}
\min_{j\in\{1,...,k\}}\{\mathbf{X}_j\}\geq (1+\eta)F. \label{eq:condition}
\end{align}
Since $N\gg 1$, the binomial distribution can be approximated by the normal distribution $\mathcal{N}(\bar{\mu},\bar{\sigma}^2)$, where $\bar{\mu}=N(1-p_1)$ and $\bar{\sigma}^2=Np_1(1-p_1)$. The statistical mean of $\min\{\mathbf{X}_j\}$ can be approximated as \cite{Royston82}
\begin{align}
\mathbb{E}\left[\min_{j\in\{1,...,k\}}\{\mathbf{X}_j\}\right]\approx \bar{\mu}+\bar{\sigma}\Phi^{-1}\left(\frac{0.625}{k+0.25}\right), \label{eq:approx}
\end{align}
where $\Phi(\cdot)$ is the cumulative distribution function (cdf) of the standard normal distribution $\mathcal{N}(0,1)$. By substituting \eqref{eq:approx} into \eqref{eq:condition}, we can solve for $N$ as
\begin{align}
N=\left\lceil\frac{2F'+p_1\beta^2+\sqrt{4p_1\beta^2F'+p_1\beta^4}}{2(1-p_1)}\right\rceil, \label{eq:single}
\end{align}
where $F'=(1+\eta)F$ and $\beta=\Phi^{-1}\left(\frac{0.625}{k+0.25}\right)$.

\begin{figure}[htb]
\centering
\includegraphics[scale=0.5]{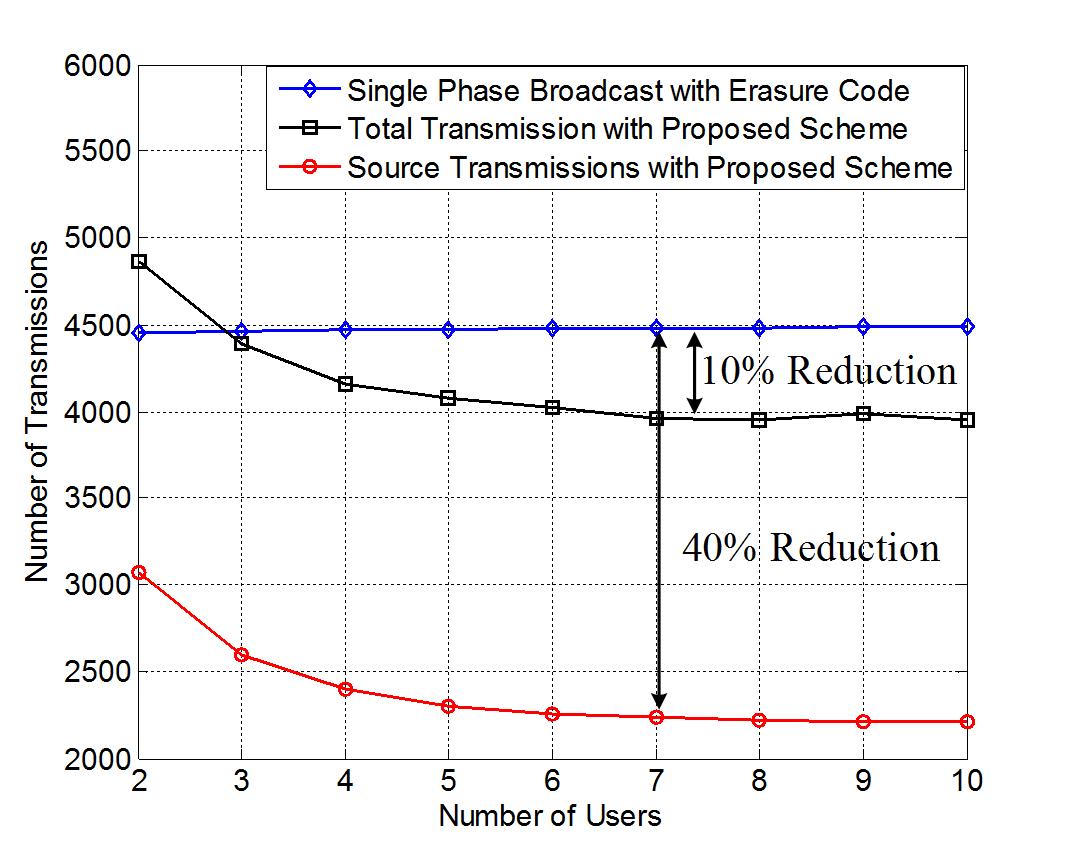}
\caption{Comparison of the proposed two-phase transmission with the single-phase transmission. }
\label{F:Compare}
\end{figure}

The performance of the proposed scheme is compared with that of the single-phase transmission, obtained in \eqref{eq:single}, for $p_1=0.5$ and $p_2=0.1$. A file containing $F=2083$ packets is to be distributed by the source node to a group of $k$ users. The coding overhead $\eta$ is set to be $5\%$ in both schemes. It is observed from Fig.~\ref{F:Compare} that the total number of transmissions increases with the number of users in the single phase scheme. However, with proposed two-phase scheme, it decreases with the number of users due to path diversity gain. In general, when $k\geq 3$,  less number of transmissions are required with the proposed scheme than that with the single-phase scheme. This is because the proposed scheme makes use of the inter-user channels in the second phase, which is more reliable than the channel between the source and the users. Furthermore, due to the long distance between the source node and the users, much higher power will be consumed by the source to compensate the path loss, as compared with the P2P transmissions. The simulation results show that the proposed scheme save more than $40\%$ of source transmissions and hence it is much more efficient in terms of power consumption at the source node.

\section{Conclusion}
In this paper, we have proposed a two-phase transmission protocol based on BATS code to achieve reliable broadcast from the source node to a group of users. With the proposed scheme, a limited number of transmissions are needed by the  source node to save its power. Furthermore, the total number of retransmissions is also slightly reduced since the inter-user channels used in Phase 2 are more reliable than the channel from the source to the users.  It is observed that the total number of transmissions may be further reduced if we slightly increase the number of source transmissions. Hence, the proposed scheme can also be designed to minimize the total number of transmissions, which will be part of our future work.

\section*{Acknowledgement}
This work was  supported by the Advanced Communications Research Program DSOCL06271, a research grant from the Directorate of Research and Technology (DRTech), Ministry of Defence, Singapore.

\bibliographystyle{ieeetr}
\bibliography{BATS}

\end{document}